\documentclass{moriond}

\def\be{\begin{equation}}
\def\ee{\end{equation}}
\def\bea{\begin{eqnarray}}
\def\eea{\end{eqnarray}}

\newcommand{\Photo}{\includegraphics[height=35mm]{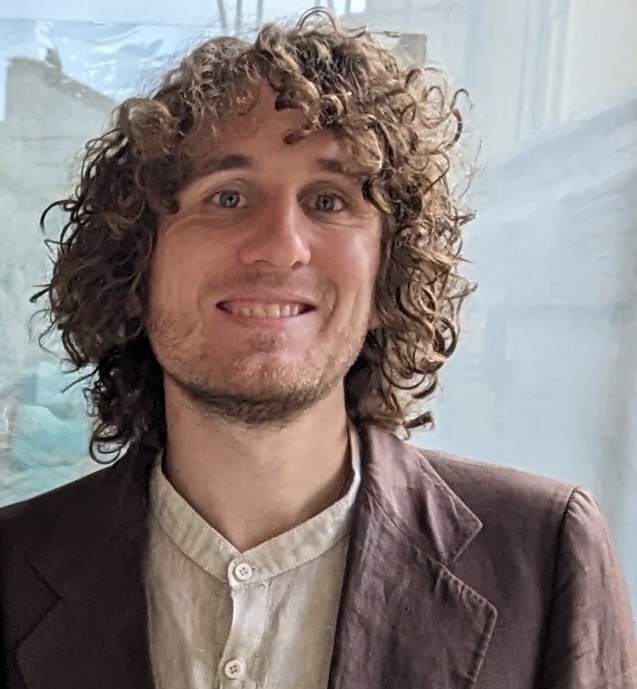}}

\usepackage{hyperref}
\usepackage{amssymb}

\begin{document}
\vspace*{4cm}

\title{Measurements of electroweak penguins and $B$ decays to final states with missing energy at Belle and Belle~II}

\author{Valerio Bertacchi, on behalf of Belle~II Collaboration }

\address{Istituto Nazionale di Fisica Nucleare (INFN), Division of Pisa
\\Largo B. Pontecorvo 3, 56127 Pisa, Italy}

\maketitle
\abstracts{The Belle and Belle II experiments have collected a 1.3 ab$^{-1}$ sample of $e^+e^-\to B\bar B$ collisions at $\Upsilon(4S)$ centre-of-mass energy. This is ideal environment to search for rare electroweak penguin $B$ decays and notably those involving $B$ decays to final states with missing energy.  Results on these datasets of $b\to s \ell^+\ell^-$ $(\ell=e,\mu)$, $b\to s\tau^+\tau^-$, and $b\to s\nu\bar \nu$ transitions are presented.}

\section{Introduction}

Belle~II~\cite{Belle2:TDR} is the experiment of the SuperKEKB collider~\cite{SuperKEKB:TDR} at KEK, Japan. SuperKEKB is an asymmetric energy $e^+e^-$ collider, which operates at centre-of-mass energy of 10.58~GeV, corresponding to the mass of the $\Upsilon(4S)$ resonance. 
The $\Upsilon(4S)$ decays almost exclusively in $B\bar B$ pairs. Belle II between 2019 and 2022 (Run 1) collected about $400\times 10^6~B\bar B$ pairs, equivalent to a integrated luminosity of $365~\mathrm{fb}^{-1}$. In the 2024 the Run 2 started and about $270~\mathrm{fb}^{-1}$ have been collected so far. The Run~2 dataset is not used in the analyses described in these proceedings. The Belle~II data sample can be combined with the $711~\mathrm{fb}^{-1}$ sample of Belle experiment, to obtain a $1.1~\mathrm{ab}^{-1}$ sample.  

The Flavour-changing neutral currents (FCNC) are processes forbidden at tree level in the Standard Model (SM) and thus suppressed. The theoretical predictions in this sector are relatively precise.  Therefore, these processes can serve as a probe to test the SM and potential access to New Physics (NP) if deviation from the SM predictions are observed.  The Belle~II clean environment and the well-known initial state produce ideal conditions to study rare $B$ decays with missing energy in the final state. These decays are reconstructed using the $B$-tagging approach. In the exclusive $B$ tagging one of the two $B$-meson ($B_\mathrm{tag}$) is reconstructed in well known hadronic or semileptonic decay channels. Then, the $B_\mathrm{tag}$ and the knowledge of $\Upsilon(4S)$ initial state are used to infer information (kinematics, flavour, charge) on the second $B$-meson of the event which is reconstructed as the required signal ($B_\mathrm{sig}$). This allows to recover the missing information from the neutrinos in the signal side.

\section{Sum-of-exclusive $B\to X_s \ell^+\ell^-$ decays}

The differential spectra of exclusive $B\to K^{(*)}\ell^+\ell^-$, $B_s\to \phi\ell^+\ell^-$ ($\ell=e,\mu$) decays have been measured by LHCb, showing $2-4\sigma$ tensions with the SM prediction in the $1<q^2<6~\mathrm{GeV}^2$ region, where $q^2$ is the squared di-lepton mass~\cite{LHCb:Kstll1,LHCb:Kstll2,LHCb:Kstll3,LHCb:Kstll-arxiv}. 
However, the inclusive spectra have smaller theoretical uncertainties ($\lesssim 7\%$)~\cite{Theory:Xsll} and they have been explored only by Belle~\cite{Belle:Xsll} and BaBar~\cite{Babar:Xsll} showing results consistent with the SM. Therefore further measurements of inclusive $B\to X_s \ell^+\ell^-$ decays are crucial to clarify the current tensions.

The analysis for $B\to X_s\ell^+\ell^-$ decays is performed on the Belle~II dataset. The signal candidate are reconstructed as the combination of two opposite-charged leptons and the $X_s$ component. The $X_s$ component is identified as the sum of 20 exclusive modes ($X_s=K,K^*$, $3K$, $Kn\pi$ with $n\leq 4, \pi^0\leq 1$). These decays cover $f_v=71\%$ of the inclusive branching fraction ($\mathcal B$), according to Monte Carlo (MC) simulation. The differential $\mathcal B$ is extracted as a function of  $q^2$ or invariant mass $M(X_s)$. 

The signal yield is extracted with a parameteric fit to the $M_\mathrm{bc}=\sqrt{(\sqrt{s}/2)^2-\vec p_B^2}$ distribution\footnote{We use natural units $\hbar=c=1$ and charge conjugation is implied throughout.}, in bin of $q^2$ or $M(X_s)$. The yield is corrected by the visible mode fraction $f_v$ and the efficiency, obtained from simulation. The signal modelling is one of the major improvement compared to previous measurements: the $\mathcal B$ of the exclusive components is updated; the description of the resonances above $1.1~\mathrm{GeV}$ is improved; the fragmatation of the inclusive region $1.1~\mathrm{GeV}$ is calibrated on $B\to X_s\gamma$ and $B\to X_sJ\psi$ data. This approach leds to a factor 3 (2) reduction of signal modelling systematic uncertainties compared to Belle (BaBar) measurements.

The backgrounds are mostly from semileptonic $B$ and $D$ decays, $B\to X_sh^+h^-$ where the hadrons are misidentified as leptons, and $B\to X_s J/\psi(\psi(2S))$ escaping the charmonium vetoes. They are suppressed using a Feature Tokenizer Transformer multivariate classifier. 

The differential $d\mathcal B/dq^2$ is reported in Fig.~\ref{fig:plots} (left) showing a good agreement with the SM prediction. In particular in $1<q^2<6~\mathrm{GeV}^2$ region $\mathcal{B}_{1-6}(B \to X_s \ell^+ \ell^-) = (1.36 \pm 0.23^{+0.13}_{-0.10}) \times 10^{-6}$, while the inclusive $\mathcal{B}(B \to X_s \ell^+ \ell^-) = (3.91 \pm 0.49^{+0.34}_{-0.26}) \times 10^{-6}$. In addition, $R_{X_s}=\mathcal B(B\to X_s\mu^+\mu^-)/\mathcal B(B\to X_se^+e^-)=0.74\pm0.19\pm0.04$ is measured for the first time, showing good agreement with the SM. 
The presented results are preliminary and currently no reference is available.

\section{$b\to s\tau^+\tau^-$ transitions: $B\to K^+\tau^+\tau^-$ and $B\to K_S^0\tau^+\tau^-$ searches}

Several NP models exhibit differently coupling with the three generations, often with an enhancement with the third one. 
Thus, $\tau$ lepton serves as a unique probe for testing these models. 
The $b\to s\tau^-\tau^+$ transitions are allowed in the SM but strongly suppressed with branching fractions~\cite{Theory:Ktautau} $\mathcal B\sim10^{-7}$. They are beyond the current experimental reach, however  several NP models predict enhancements of these transitions. For instance NP models which accommodate lepton flavour universality tensions or $B^+\to K^+\nu\bar\nu$ decays excess, also increase the $\mathcal B(B\to K^{(*)}\tau^+\tau^-)$ of several orders of magnitude~\cite{Theory:Ktautau}. 

The search of $B\to K^+\tau^+\tau^-$ decays is performed on the combined Belle and Belle~II dataset, using the hadronic $B$-tagging method. The signal is reconstructed using the leptonic $\tau$ decays only. The background is reduced applying consecutive cuts, without the usage of MVA classifiers. The most important selection variable is the invariant mass $m(K^+\ell^-)$ which is required to be larger than 1.9~GeV to suppress th dominant $B\to D^{(*)}\ell\nu$ background. The background is further reduced optimizing the selection on lepton $\ell^+$ momentum, missing mass and $E_\mathrm{extra}$ (the sum of the calorimetric energy of the photons not associated with a track). The residual background is calibrated on sidebands. The signal is extracted counting the events in the first bin of $E_\mathrm{extra}$, corresponding to $E_\mathrm{extra}<0.1~(0.25)$~GeV for Belle (Belle~II) sample. No significant signal is observed, thus an an Upper Limit (UL) is set: $\mathcal B(B\to K^+\tau^+\tau^-)<0.56\times 10^{-3}$ at 90\%~CL. This is the world best limit for the channel, improving by a factor four the previous best limit~\cite{Babar:KpTauTau}. The presented results\footnote{This UL is different from the conference value as a result of an update of the derivation during the review.} are preliminary and currently available on Ref.~\cite{Belle2:KpTauTau}. 

\begin{figure}[!hbt]
  \centering
  \begin{minipage}[c]{0.48\textwidth}
    \includegraphics[height= 5.2cm]{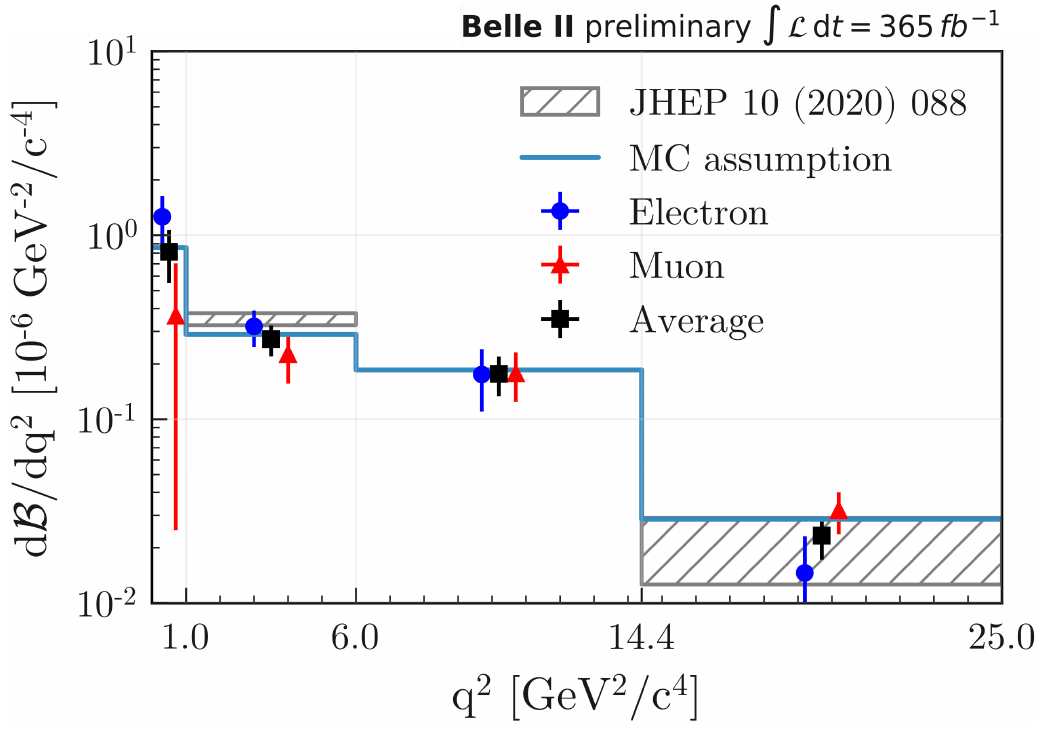}
  \end{minipage}
  \hspace{0.4cm}
  \begin{minipage}[c]{0.48\textwidth}
    \centering
    \includegraphics[height=5 cm]{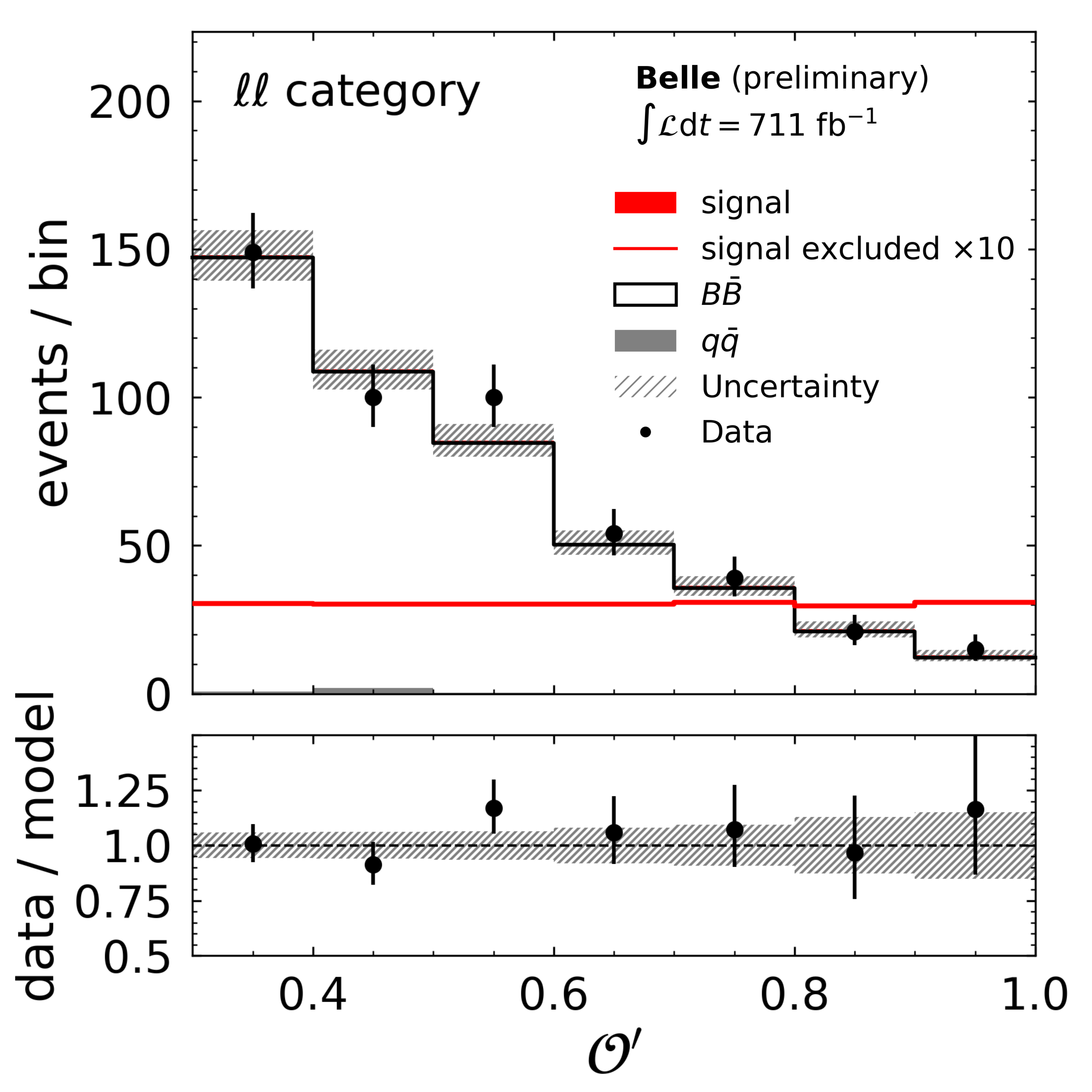}
  \end{minipage}
\caption{Left:$d\mathcal B(B\to X_s\ell^+\ell^-)/dq^2$ in $\ell=e,\mu$, and combined channel; the SM prediction and the MC distribution assuming the inclusive $\mathcal B$ are overlaid. Right: Post-fit distribution of the BDT output for $B\to K_S^0\tau^+\tau^-$ search; MC signal at $\mathcal B=8.4\times 10^{-3}$ is overlaid. }\label{fig:plots}
\end{figure}

The search of $B\to K_S^0\tau^+\tau^-$ decays is performed on the combined Belle and Belle~II dataset, using the hadronic $B$-tagging method. The signal is reconstructed targeting $\tau$ one-prong decays, reconstructing $\tau\to e\nu\bar \nu,~\mu\nu\bar\nu,~\pi\nu,~\rho(\to \pi\pi^0) \nu$ channels. The reconstrcted events are subdivided in five categories according to the final state of the two $\tau$'s ($\ell\ell$, $\ell h$, $h\ell$, $\rho \ell$ and no-$\ell$). The background is suppressed with a Boosted Decision Tree (BDT) based on missing energy, $E_\mathrm{extra}$, $q^2=(p_{\tau^+}+p_{\tau^-})^2$ and additional kinematic variables. The residual background is calibrated on sidebands. The signal is extracted with a binned maximum likelihood fit to the BDT output distributions of the 10 categories (5 for each dataset). As an example, in Fig.~\ref{fig:plots} (right) is shown the distribution of the BDT output for the signal region in $\ell\ell$ category for Belle dataset. No significant signal is observed, thus an UL is set: $\mathcal B(B\to K_S^0\tau^+\tau^-)<0.84\times 10^{-3}$ at 90\%~CL. This is the first search of this channel. The presented results are preliminary and currently no reference is available.

\section{Search for $B\to X_s\nu\bar\nu$ decays}

The inclusive $\mathcal B(B\to X_s\nu\bar\nu)=(3.48\pm0.11)\times 10^{-5}$ in the SM~\cite{Theory:Xsnunu}. The best UL for this branching fraction has been set by ALEPH collaboration at $\mathcal B(b\to s\nu\bar\nu)<6.4\times 10^{-4}$ level~\cite{ALEPH:Xsnunu}. This analysis is part of the effort of Belle~II of in-depth investigation of the  $b\to s \nu\bar\nu$ sector, following the tension created by the evidence $B^+\to K^+\nu\bar \nu$ decays~\cite{Belle2:Knunu}. In particular, the inclusive branching fraction is sensitive to different NP parameters compared to the exclusive $B\to K^{(*)}\nu\bar \nu$~\cite{Theory:Knunu_inclusive}. 

The analysis is performed on the Belle~II data sample using the hadronic $B$-tagging method. The signal is identified as the sum of 30 exclusive modes ($B\to Kn\pi$ with $n\leq 4$, $B\to 3K$, $B\to 3K\pi$). 
The backgrounds are suppressed with a BDT selection, and the signal is extracted with a binned maximum likelihood fit to the bidimensional distribution of BDT score $O(BDT)$ and the invariant mass of the $X_s$ system ($M(X_s)^\mathrm{reco}$). The $q\bar q$ continuum background estimation is validated using the off-resonance sample. The $B\overline B$ background is validated using a sideband reverting the $O(BDT)$ selection and using an $M_{bc}$ sideband.

No significant signal is observed, thus ULs in bin of $M(X_s)$ are set. At 90\% CL $\mathcal B(B\to X_s\nu\bar\nu)< 2.2 \times 10^{-5}~(0<M(X_s)<0.6~\mathrm{GeV}),~<9.5\times 10^{-4}~(0.6<M(X_s)<1.0~\mathrm{GeV}),~3.12\times 10^{-4}~(1~\mathrm{GeV}<M(X_s))$. 
 Combining all the invariant mass regions we obtain the UL $\mathcal B(B\to X_s\nu\bar\nu)<3.2\times 10^{4}$. This results are new world's best UL for this channel. The lowest $M(X_s)$ limit is compatible with the hadronic-tagged measurement of the exclusive $\mathcal B(B^+\to K^+\nu\bar \nu)$~\cite{Belle2:Knunu}. The presented results are preliminary and currently available on Ref.~\cite{Belle2:Xsnunu}.

\section{$B^+\to K^+\nu\bar\nu$ reinterpretation}

The first evidence of $B^+\to K^+\nu\bar \nu$ decays~\cite{Belle2:Knunu} has a $2.7\sigma$ tension with the SM, with a measured branching fraction about four times the expected one. This result received significant interest in the theory community, which provided several possible interpretations of the excess~\cite{Theory:Knunu_LFV,Theory:Knunu_NP_1,Theory:Knunu_NP_2}. 

To facilitate the use of the Belle~II results in a complete and correct way, a novel reinterpretation method has been developed~\cite{Gartner:reinterpretation}. The goal is to build a model-agnostic likelihood and reweight it to the desired model. Assuming $q^2$ to be the relevant kinematic variable of the model, and $x$ to be the fitting variable, the central quantity is the number density
$n(x)=L\int\varepsilon (x|q^2)\sigma(q^2) dq^2$, where $L$ is the luminosity and $\varepsilon$ the efficiency. Given a reference number density $n_0(x)$, like the SM one, we can obtain an alternative model number density reweighting it: $n_1(x)=\sum_{q^2~\mathrm{bins}}n_{0,q^2}(x)w(q^2)$, with $w(q^2)=\sigma_1(q^2)/\sigma_0(q^2)$.

We applied this method to the $B^+\to K^+\nu\bar \nu$ analysis~\cite{Belle2:Knunu-reinterpretation}. Here the fitting variable is $x=\eta(BDT2)\times q_\mathrm{rec}^2$ (the output of the selection BDT and the squared di-neutrino mass $q^2=(p_B-p_K)^2)$, while $n_0(x)$ is the SM signal. The result is reinterpreted in term of Weak Effective Theory (WET) including dimension 6 operators. In this framework we introduce left (L) and right (R) scalar (S), vector (V) and tensor (T) Wilson coefficients $C$, while in the SM only the $C_\mathrm{VL}$ coefficient is different from 0. 

The WET signal has a significance of $3.3\sigma$ versus the background hypothesis only. Moreover, a non-zero tensor contribution and a larger vector contribution are preferred compared to the SM expectation. 
In addition, highest density credible intervals (the smallest possible credible intervals at a given probability) are provided. 
In conclusion, the main deliverables of this analysis are the likelihood~\cite{Belle2:Knunu-reinterpretation-hepdata} and the joint number density distribution of $B^+\to K^+\nu\bar \nu$, together with the method and the tools to facilitate further reinterpretations. 

\section*{References}
\bibliography{biblio.bib}

\end{document}